\documentclass[twocolumn,superscriptaddress,showpacs,preprintnumbers,amsmath,amssymb,floatfix]{revtex4}

\usepackage{graphicx,amssymb}

\begin{document}

\title{Lane-formation vs. cluster-formation in two dimensional
    square-shoulder systems: \\ A genetic algorithm approach}

\author{Julia Fornleitner} 
\email{fornleitner@cmt.tuwien.ac.at}
\affiliation{Center for Computational Materials Science and Institut
f\"ur Theoretische Physik, Technische Universit\"at Wien, Wiedner
Hauptstra{\ss}e 8-10, A-1040 Wien, Austria} 

\author{Gerhard Kahl}
\affiliation{Center for Computational Materials Science and Institut
f\"ur Theoretische Physik, Technische Universit\"at Wien, Wiedner
Hauptstra{\ss}e 8-10, A-1040 Wien, Austria}

\date{\today}

\begin{abstract}
Introducing genetic algorithms as a reliable and efficient tool to
find ordered equilibrium structures, we predict minimum energy
configurations of the square shoulder system for different values of
corona width $\lambda$. Varying systematically the pressure for
different values of $\lambda$ we obtain
complete sequences of minimum energy configurations which
provide a deeper understanding of the system's strategies to arrange
particles in an energetically optimized fashion, leading to the
competing self-assembly scenarios of cluster-formation
vs. lane-formation.
\end{abstract}

\maketitle

The ability of colloidal dispersions to self-organize in a
surprisingly large variety of ordered structures markedly
distinguishes soft matter from hard (i.e., atomic) materials. The
spectrum of ordered particle arrangements encountered in soft matter
systems does not only contain low-symmetry, non-close-packed
structures, such as bco, diamond, or the A15 lattices
\cite{Zih00,Zih01,Wat99,Got04,Got05a}; soft matter particles are able
to self-organize in considerably more complex ways, forming thereby
micellar and inverse micellar structures \cite{Pie06,Gla07}, cluster
phases \cite{Mla06}, chain-like or layered arrangements
\cite{Jag98,Jag99a,Cam03,Mal03,Mal04,Gla07}, or gyroid phases
\cite{Haj94,Mat98,Ull04}, just to name a few of them.

In the investigation of self-organizing phenomena in soft matter
systems, theoreticians have developed reliable tools.
On one side efficient and accurate coarse-graining procedures have
been developed which, by averaging over the large number of degrees of
freedom of the solvent particles \cite{Lik01,Bal04}, lead to effective
potentials between the colloidal particles. On the other side there
are statistical mechanics based concepts which provide on the basis of
these interactions reliable information on the thermodynamic
properties that finally lead to the phase diagram of the system at
hand. Among those are density functional theory \cite{Eva92}, liquid
state theories \cite{Han06}, or computer simulations \cite{Fre02}.

What is badly missing among the theoreticians' tools is a reliable way
of how to {\it predict} the ordered equilibrium structures for a given
system. In hard materials one can safely rely on experience,
intuition, or plausible arguments: a set of pre-selected candidate
structures is chosen, comprising the usual suspects (such as fcc, bcc,
etc.) and at a given state point the lattice with the lowest (free)
energy is the stable one. In soft matter, however, the situation is
entirely different: as a consequence of the rich wealth of emerging
equilibrium structures a conventional approach that is based on a
biased pre-selection process is bound to fail. Computer simulations
are not very helpful either, since they are time-consuming and risk to
be trapped in local energetic minima due to the rough and complex
energy surface.

In this contribution we present an alternative strategy which we
believe to be very powerful in the search of equilibrium structures
and which does not share the deficiencies of the conventional
approaches. We use genetic algorithms (GAs), i.e., optimization
strategies, that adopt features of evolutionary processes as key
elements to find the optimal solution for a problem \cite{Hol75}. In
contrast to the conventional methods, GAs allow for searching
basically among {\it all} lattice structures in a parameter-free and
unbiased way. Despite the fact that GAs have been proposed several
decades ago \cite{Hol75}, their usefulness in problems of condensed
matter theory has been acknowledged only in recent years (see, e.g.,
\cite{Oga06,Sie07}). In the present contribution we demonstrate the
power of this approach for a two-dimensional (2D) system where
the particles interact via spherically symmetric square-shoulder
potentials. With the help of GAs we identify minimum energy
configurations (MECs), or, equivalently the zero-temperature phase
diagram.

GAs mimic certain principles and processes known from natural
evolution like mutation, mating and `survival of the fittest' to
find the global extremum in the function to optimise. Because of their
particular search strategies they are able to investigate large areas
in search space and to concentrate at the same time their efforts on
the most promising regions. It is in particular this global scope which has
made GAs a highly appreciated tool in many fields.

The basic unit of a GA is a so-called {\it individual}, ${\cal I}$,
which represents one possible solution of the problem, encoded in a
string of genes. Each gene can take on values out of a chosen
alphabet. The encoding of the candidate to the individual is a very
delicate task that has major influence on the performance of the GA. A
set of individuals is called {\it generation}.

With these definitions we can sketch the work-flow of a GA: first a
starting generation is created at random, whose individuals consist of
arbitrary sequences of genes. Each of the individuals ${\cal I}$ is
assigned a `fitness-value' through a problem-specific function
$f({\cal I})$, assessing the quality of the solution represented by
the individual. A higher fitness-value marks a better solution in this
evaluation process. According to their fitness, parent individuals are
selected for reproduction to generate new individuals by simple
operations like recombination and mutation, for which many different
methods have been proposed in literature.  This circle, i.e.,
evaluation--selection--recombination--mutation, is iterated until a
sufficiently large number of generations has been created. The
individual with the absolutely highest fitness value is taken as the
final solution. Since GAs do not converge exactly to the global
minimum, additional refinement of the proposed solution is required.

In an effort to adapt this general algorithm to our particular
problem, we first have to find a suitable encoding of the candidate
structures to the individuals. To this end we have utilized two
different strategies: {\bf (i)} a simple lattice parametrisation where
the lattice vectors and positions of eventual basis particles are
translated via the binary alphabet to strings of 0s and 1s; {\bf (ii)}
a more refined strategy, which is specially adapted to systems for
which cluster phases are to be expected, has been developed: in this
`cluster-biased' version the lattice sites are populated by ordered
(i.e. regular) dimers or trimers. Hence individuals include additional
information describing the orientation and the distance of particles
in the cluster. In this way more particles per unit cell can be
considered, reducing at the same time the number of parameters. This
leads to a considerably enhanced performance of the GA for complex
lattices. In addition, for hard-core particles, particular care is in
order to prevent the creation of individuals that represent unphysical
configurations with overlapping cores.

Working at fixed particle number $N$, pressure $P$, and temperature
$T$, we have decided for the standard form of $f({\cal I})$, i.e.,
\begin{equation}
f({\cal I}) = \exp \{-[G({\cal I}) - G_0]/G_0 \} \quad .
\end{equation}
At $T=0$, the Gibbs free energy, $G$, reduces to $G=U+PV$, $V$ being
the volume. $G_0$ stands for the Gibbs free energy of a reference
structure.  For further conceptual and numerical details concerning
the GA we refer to \cite{Got05b,For07}.

The interaction potential of the square-shoulder system consists of an
impenetrable core of diameter $\sigma$ with an adjacent step-shaped,
repulsive corona (with range $\lambda \sigma$), i.e.,
\begin{equation}
\Phi(r) = \left\{ \begin{array} {l@{~~~~~~~~}l} 
                  \infty & r \le \sigma \\ 
                  \epsilon & \sigma < r < \lambda \sigma \\
                             \end{array}
                             \right. ,
\end{equation}
$\epsilon$ being the height of the shoulder. The search for MECs for
this particular system represents undoubtedly the most stringent test
for the GA: with its flat plateau and as sharp a cutoff as possible the MECs
can be easily classified by the number of overlapping coronas which makes the potential a quintessential
\cite{Zih01} test system.  The fact that we restrict ourselves to a 2D
system does not represent a limitation at all. On the contrary, the
particle arrangements can be visualized in a very convenient way that
makes it much easier to understand the sequence of MECs as the
pressure is increased. Objections that the system is oversimplified
and only of academic interest can easily be refuted: there exist a
number of realistic soft systems with a core-corona-architecture that
can be described using a square-shoulder interaction
\cite{Mal04}. Among these are, for instance, colloidal particles with
block-copolymers grafted to their surface where self-consistent field
calculations lead to effective interactions that closely resemble
$\Phi(r)$ given in (2) \cite{Nor05}.

The search for MECs for the square-shoulder potential has been
pioneered by Jagla \cite{Jag98,Jag99a} and was carried on in later
work by Malescio and Pellicane \cite{Mal03,Mal04} using Monte Carlo
simulations and geometrical considerations. Further theoretical work
on this system was presented in \cite{Gla07}. The results are indeed
remarkable: a large variety of MECs has been identified where --
despite the radial symmetry of $\Phi(r)$ -- particles often arrange in
asymmetric structures, forming thereby lanes or ring-like
structures. However, the authors of above mentioned studies raise
doubts themselves that their set of MECs might not be
complete. Introducing GAs as a novel technique to this particular
problem, we will present in the following sequences of MECs that show
a considerably larger variety than the ones identified up to now and
which can easily be understood via energetic arguments. Although we
cannot provide a rigorous proof there is evidence that these sequences
are complete.

Our results offer -- depending on the range of the shoulder $\lambda$
-- a deeper insight that the square-shoulder system develops different
strategies to form MECs. Among the systems we have investigated we
present in the following results for three different values for
$\lambda$, corresponding to a small ($\lambda=1.5$), an intermediate
($\lambda=4.5$), and a large shoulder range ($\lambda=10$). Standard
reduced units are used: $P^\star=P\sigma^2/\epsilon$,
$U^\star=U/(N\epsilon)$ and $G^\star=G/(N\epsilon) =
U^\star+P^\star/(\eta\sigma^2)$, $\eta = N/V$ being the number
density.

\begin{figure}
      \begin{center}
      \begin{minipage}[t]{8.5cm}
      \includegraphics[width=8.4cm, clip]
{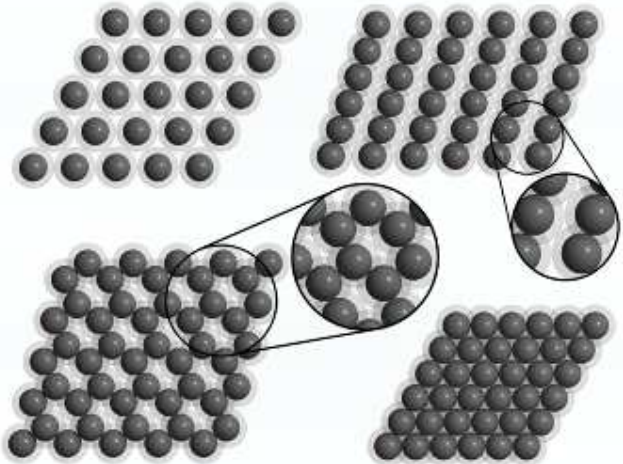}
      \end{minipage}
      \end{center}
\caption{MECs for the square-shoulder system of shoulder range
$\lambda = 1.5$. Configurations correspond (from left to right and
from top to bottom) to pressure values indicated in
Fig. \ref{fig:thd_1.5} by vertical arrows.}
\label{fig:str_1.5}
\end{figure}

\begin{figure}
      \begin{center}
	\begin{minipage}[t]{6.5cm}
      \includegraphics[width=6.4cm, clip]
{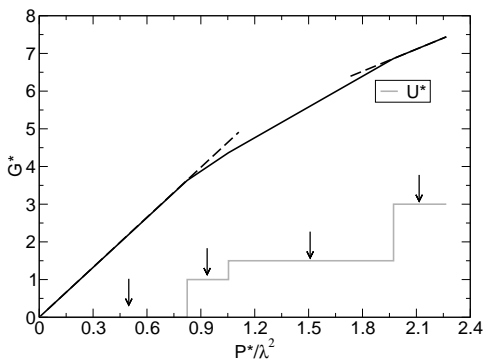}
      \end{minipage}
      \end{center}
\caption{$G^\star$ (black line) and $U^\star$ (grey line) as functions
of $P^\star/\lambda^3$ for the square-shoulder system with $\lambda =
1.5$. Vertical arrows indicate MECs depicted in
Fig. \ref{fig:str_1.5}. Broken lines represent limiting cases of MECs
(see text).}
\label{fig:thd_1.5}
\end{figure}

We start with $\lambda = 1.5$ and show in Fig. \ref{fig:str_1.5} the
MECs proposed by the GA, while Fig. \ref{fig:thd_1.5} displays the
corresponding thermodynamic properties $G^\star$ and $U^\star$ as
functions of $P^\star$. At very low pressure, particles populate an ideal
 hexagonal lattice, thus avoiding overlapping coronas. Upon
compression, the system must pay in some form tribute to the reduced
space in terms of an energy penalty, i.e. via a first overlap of
shoulders. Obviously, for this $\lambda$-value, the formation of lanes is energetically the best
solution. Along these lanes, particles are in direct contact (and,
consequently have overlapping coronas), forming a one-dimensional
close-packed arrangement.  Parallel lanes, however, try to avoid
corona overlap and the shoulder width $\lambda$ serves as a spacer
(see magnified view). As the pressure is further increased, new
strategies are required. While particles still prefer alignment along
lanes their internal arrangement is modified: rather than forming
straight lines, the lanes are zig-zag shaped due to energetic reasons,
which is a compromise between the reduced available space and the
imminent energetic penalty due to additional corona
overlaps. Neighbouring lanes are arranged in such a way, that each
particle is now in direct contact with three other
ones. Alternatively, the staggered lanes can also be viewed as a
ring-like structure: six particles form elongated rings where
$\lambda$ serves again as a spacer, fixing the width of the
cage. In the end, further compression causes the system to collapse
into the close-packed hexagonal structure where each particle is in
direct contact with six neighbours.


The functional form of $G^\star$ and $U^\star$ can nicely be understood via the following,
general thermodynamic considerations (cf. Fig. \ref{fig:thd_1.5}):
For a given particle configuration, characterized by the number of
overlapping coronas, $U^\star$ is constant; thus $G^\star = U^\star +
P^\star/(\eta \sigma^2)$ is a straight line as a function of $P^\star$
with slope $1 / \eta$. Two limiting cases (indicated as broken lines
in Fig. \ref{fig:thd_1.5} and characterized by slopes $1/\eta_{\rm
min}$ and $1/\eta_{\rm max}$) can easily be identified: an arrangement
where disks of diameter $\lambda$ form a close-packed structure (with
$\eta_{\rm min}$ and $U^\star = 0$) and the hexagonal lattice where
the particles' hard cores form a close-packed structure (with $\eta_{\rm max}$ and
$U^\star = U^\star_{\rm max} = 3$). All other MECs identified by the
GA are located on lines of slope $1/\eta$, with $1/\eta_{\rm max} \le
1/\eta \le 1/\eta_{\rm min}$, and obviously have lower
$G^\star$-values.  Thus $G^\star = G^\star(P^\star)$ becomes a
sequence of intersecting straight lines, representing a considerable
help in analysing the data. Cross-over points between two competing structures
where GAs tend to have convergence problems \cite{Got05b} can be
determined {\it exactly} as the intersection of two straight
lines. The energy levels $U^\star$ (also depicted in Fig.~\ref{fig:thd_1.5}) that characterize the MECs
are of course rational numbers, i.e., number of overlapping coronas
divided by the number of particles per unit cell. Finally, all calculations have been performed on rather fine
pressure grids ($\Delta P^\star \sim 0.2$) so we are confident
that the presented sequences of MECs are complete.

\begin{figure}
      \begin{center}
      \begin{minipage}[t]{8.5cm}
      \includegraphics[width=8.4cm, clip]
{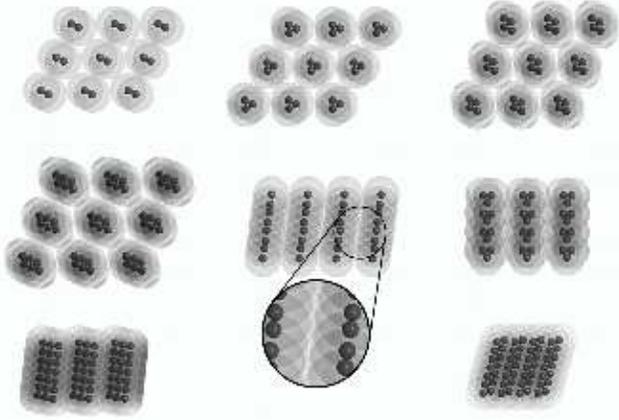}
      \end{minipage}
      \end{center}
\caption{MECs for the square-shoulder system of shoulder range
$\lambda = 4.5$. Configurations correspond (from left to right and
from top to bottom) to pressure values indicated in
Fig. \ref{fig:thd_4.5} by vertical arrows.}
\label{fig:str_4.5}
\end{figure}

\begin{figure}
      \begin{center}
      \begin{minipage}[t]{6.5cm}
      \includegraphics[width=6.4cm, clip]
{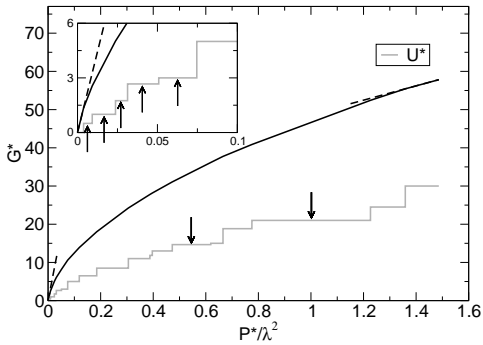}
      \end{minipage}
      \end{center}
\caption{$G^\star$ (black line) and $U^\star$ (grey line) as functions
of $P^\star/\lambda^3$ for the square-shoulder system with $\lambda =
4.5$. Vertical arrows indicate MECs depicted in
Fig. \ref{fig:str_4.5}. Broken lines represent limiting cases of MECs
(see text).}
\label{fig:thd_4.5}
\end{figure}

As we proceed to $\lambda = 4.5$ the systems develops completely
different strategies to form MECs as the pressure is increased; they
are depicted in Fig.  \ref{fig:str_4.5}, while Fig. \ref{fig:thd_4.5}
displays the thermodynamic properties. The hexagonal pattern imposed
by the non-overlapping coronas (observed at extremely low
pressure-values and not displayed here) is soon superseded by a novel
strategy, namely cluster formation. At low pressure, dimers are formed
which populate the sites of a distorted hexagonal lattice. Upon
further compression, these aggregates become larger until they reach
the size of six particles. The degree of distortion of the underlying
hexagonal lattice is imposed by the shape of the clusters: therefore,
the trimers, which have nearly circular shape, sit on a nearly perfect
hexagonal lattice, while for elongated hexamers the structure is
strongly distorted. As the system is further compressed, formation of
clusters is obviously energetically less attractive and lane-formation
sets in. With increasing pressure the structure of the MECs becomes
more complex. In the beginning each lane is built up by a linear
sequence of dimers and thus closely resembles the first lane-scenario
observed for $\lambda = 1.5$. At higher pressure-values, however, the
system forms parallel lanes which are now built up by larger
clusters. The increasing complexity of this inner structure makes
simple energetic explanations in terms of overlapping coronas
impossible. However, the grey shades in Fig. \ref{fig:str_4.5} give
evidence that the formation of clustered lanes is an efficient
strategy to avoid an overlap between neighbouring lanes, which
represents a considerably higher energy penalty. The tendency to form
parallel lanes is maintained until the system finally collapses into
the close-packed hexagonal structure. The considerably richer wealth
of MECs encountered for $\lambda = 4.5$ is reflected by the large
number of energy levels in the plot $U^\star$ vs. $P^\star$ and the
number of intersecting straight line segments in the diagram of
$G^\star$ vs. $P^\star$.


\begin{figure}
      \begin{center}
      \begin{minipage}[t]{8.5cm}
      \includegraphics[width=8.4cm, clip]
{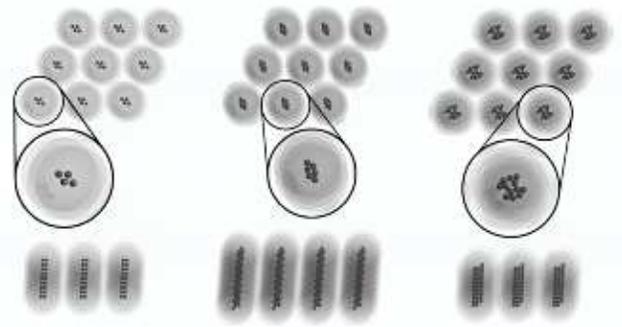}
      \end{minipage}
      \end{center}
\caption{MECs for the square-shoulder system of shoulder range
$\lambda = 10$. Configurations correspond (from left to right and
from top to bottom) to pressure values indicated in
Fig. \ref{fig:thd_10} by vertical arrows.}
\label{fig:str_10}
\end{figure}

Finally, for $\lambda = 10$, the strategies of the system to form MECs
seem at first sight similar to the previous case: again, the system
prefers formation of clusters at low pressure, which are located on
slightly distorted hexagonal lattices. However, we observe that inside
a cluster the cores of the particles are sometimes arranged in a
disordered fashion, while for $\lambda = 4.5$ only clusters with an ordered
internal particle arrangement occur.  The strategy is obviously the following: once $\lambda$ is sufficiently large so
that cluster formation is supported, the system tries to arrange
particles so that the shape of the cluster becomes as circular as
possible. This in turn guarantees that the underlying structure is
close to the energetically most favourable hexagonal lattice. For
$\lambda = 4.5$, where the core region still represents a considerable
fraction of the particle diameter, the system has to proceed rather
carefully to fulfill this requirement, leading to the ordered
arrangements of the cores. For $\lambda = 10$, however, the core
region is nearly negligible with respect to the core width. Now both
regular and irregular particle arrangements inside the core can lead
to circular-shaped clusters of the same size, having practically the same $G$-value.  Again, at higher pressure values
lanes with an increasingly complex inner structure are formed. Some of
them (for example, the last one displayed in Fig. \ref{fig:str_10})
might be comparable to the Bernal spirals observed experimentally in
three dimensional colloidal systems \cite{Cam05}. Lane formation
persists until the system finally collapses into the close-packed
hexagonal structure. If the square-shoulder particles still represent a reasonable model for
macromolecules at such high pressure values is questionable.

\begin{figure}
      \begin{center}
      \begin{minipage}[t]{6.5cm}
      \includegraphics[width=6.4cm, clip]
{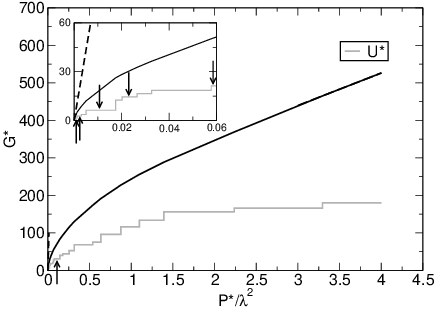}
      \end{minipage}
      \end{center}
\caption{$G^\star$ (black line) and $U^\star$ (grey line) as functions
of $P^\star/\lambda^3$ for the square-shoulder system with $\lambda =
10$. Vertical arrows indicate MECs depicted in
Fig. \ref{fig:str_1.5}. Broken lines represent limiting cases of MECs
(see text).}
\label{fig:thd_10}
\end{figure}

Let us come back to the disordered clusters. If we extrapolate
these results to even larger values of $\lambda$ -- or, equivalently,
to a vanishing core -- and briefly switch to three dimensions, we arrive at another soft matter model system
that has been studied in detail in literature: the penetrable sphere
model (PSM) \cite{Lik98}. Among the remarkable, well-documented
features of this system is its ability to form cluster phases: at
sufficiently high densities clusters of overlapping particles populate
the lattice sites of a regular fcc lattice. Detailed simulations of a
closely related system that also shows clustering \cite{Mla06} have
revealed that the internal structure of the clusters is completely
random. These observations are consistent with our results in the
sense, that an increasing corona width (or, equivalently, a vanishing
core) favours formation of disordered clusters of particles, which in
turn populate sites of regular lattices. Obviously first precursors of this
phenomenon occur at $\lambda=10$ \cite{Lik07}.

Our results provide an explanation for the system's strategy to form MECs: if
$\lambda$ is small, formation of larger clusters is prohibited due to geometric
reasons, hence particles arrange in lanes of variable shape or in connected
structures. However, as soon as the range of the corona is sufficiently large,
the formation of clusters sets in and they dominate the low-pressure
regime. Depending on the corona range, the particles group in ordered or
disordered clusters, keeping the cluster's shape as disk-like as
and the underlying structure as close to the ideal hexagonal lattice as possible. For systems that exhibit clustering, lane formation sets in
at higher pressure values. The lanes themselves are built up by clusters, so one is led to
interpret the crossover to the lane regime as a structural change of the
cluster crystal.

The results of our investigations are, on one side, of rather basic
relevance. The simple form of the potential allows to understand the
full sequence of MECs on an energetic level as the pressure is
increased and thus provides deeper insight under which conditions the
system prefers to form either lanes or clusters. But also from a more
applied point of view, these results are of relevance. Having
established GAs as a suitable and reliable tool for finding ordered
equilibrium structures, the strategies of self-assembly of systems
with a substantially more complex effective interparticle interaction
can be better understood. Such knowledge is certainly of relevance in
technological applications, such as nanolithography or
nanoelectricity.

{\bf Acknowledgements} The authors are indebted to Dieter Gottwald and
Gernot J. Pauschenwein (both Wien), Christos N. Likos (D\"usseldorf),
and Primoz Ziherl (Ljubljana) for stimulating discussions. Financial
support by the Austrian Science Foundation (FWF) under
Proj. Nos. P15758-N08 and P19890-N16 is gratefully acknowledged.



\begin{thebibliography}{10}

\bibitem{Zih00}
Ziherl, P.; Kamien, R. {\em Phys. Rev. Lett.} {\bf 2000}, {\em 85}, 3528.

\bibitem{Zih01}
Ziherl, P.; Kamien, R. {\em J. Phys. Chem. B} {\bf 2001}, {\em 105}, 10147.

\bibitem{Wat99}
Watzlawek, M.; Likos, C.; L{\"o}wen, H. {\em Phys. Rev. Lett.} {\bf 1999}, {\em
  82}, 5289.

\bibitem{Got04}
Gottwald, D.; Likos, C.; Kahl, G.; L{\"o}wen, H. {\em Phys. Rev. Lett.} {\bf
  2004}, {\em 92}, 068301.

\bibitem{Got05a}
Gottwald, D.; Likos, C.; Kahl, G.; L{\"o}wen, H. {\em J. Chem. Phys.} {\bf
  2005}, {\em 122}, 074903.

\bibitem{Pie06}
Pierleoni, C.; Addison, C.; Hansen, J.-P.; Krakoviack, V. {\em Phys. Rev.
  Lett.} {\bf 2006}, {\em 96}, 128302--1.

\bibitem{Gla07}
Glaser, M.; Grason, G.; Kamien, R.; Ko$\mathrm{\check s}$mrlj, A.; Santangelo,
  C.; Ziherl, P. {\em Europhys. Lett.} {\bf 2007}, {\em 78}, 46004.

\bibitem{Mla06}
Mladek, B.; Gottwald, D.; Kahl, G.; Neumann, M.; Likos, C. {\em Phys. Rev.
  Lett.} {\bf 2006}, {\em 96}, 045701.

\bibitem{Jag98}
Jagla, E. {\em Phys. Rev. E} {\bf 1998}, {\em 58}, 1478.

\bibitem{Jag99a}
Jagla, E. {\em J. Chem. Phys.} {\bf 1999}, {\em 110}, 451.

\bibitem{Cam03}
Camp, P. {\em Phys. Rev. E} {\bf 2003}, {\em 68}, 061506.

\bibitem{Mal03}
Malescio, G.; Pellicane, G. {\em Nat. Materials} {\bf 2003}, {\em 2}, 97.

\bibitem{Mal04}
Malescio, G.; Pellicane, G. {\em Phys. Rev. E} {\bf 2004}, {\em 70}, 021202.

\bibitem{Haj94}
Hajduk, D.; Harper, P.; Gruner, S.; Honeker, C.; Kim, G.; Thomas, E.; Fetters,
  L. {\em Macromolecules} {\bf 1994}, {\em 27}, 4063.

\bibitem{Mat98}
Matsen, W. {\em J. Chem. Phys.} {\bf 1998}, {\em 108}, 785.

\bibitem{Ull04}
Ullal, C.; Maldovan, M.; Thomas, E.; Chen, G.; Han, Y.; Yang, S. {\em Phys.
  Rev. Lett.} {\bf 2004}, {\em 84}, 5434.

\bibitem{Lik01}
Likos, C. {\em Phys. Rep.} {\bf 2001}, {\em 348}, 267.

\bibitem{Bal04}
Ballauff, M.; Likos, C. {\em Angew. Chemie Intl. English Ed.} {\bf 2004}, {\em
  43}, 2998.

\bibitem{Eva92}
Evans, R.;
\newblock Marcel Dekker, New York, 1992;
\newblock chapter~3, pages 85--175.

\bibitem{Han06}
Hansen, J.-P.; McDonald, I. {\em Theory of Simple Liquids;}
\newblock Elsevier, Amsterdam, 3rd ed., 2006.

\bibitem{Fre02}
Frenkel, D.; Smit, B. {\em Understanding Molecular Simulations;}
\newblock Academic Press, London, 2nd ed., 2002.

\bibitem{Hol75}
Holland, J. {\em Adaption in Natural and Artificial Systems;}
\newblock The University of Michigan Press, Ann Arbor, 1975.

\bibitem{Oga06}
Oganov, A.; Glass, C. {\em J. Chem. Phys.} {\bf 2006}, {\em 124}, 244704.

\bibitem{Sie07}
Siepmann, P.; Martin, C.; Vancea, I.; Moriarty, P.; Krasnogor, N. {\em Nano
  Letters} {\bf 2007}, {\em 7}, 1985--1990.

\bibitem{Got05b}
Gottwald, D.; Kahl, G.; Likos, C. {\em J. Chem. Phys.} {\bf 2005}, {\em 122},
  204503--1.

\bibitem{For07}
Fornleitner, J.; Kahl, G. {\em to be published}

\bibitem{Nor05}
Norizoe, Y.; Kawakatsu, T. {\em Europhys. Lett.} {\bf 2005}, {\em 72}, 583.

\bibitem{Cam05}
Campbell, A.; Anderson, V.; van Duijneveldt, J.; Bartlett, P. {\em Phys. Rev.
  Lett.} {\bf 2005}, {\em 94}, 208301.

\bibitem{Lik98}
Likos, C.~N.; Watzlawek, M.; L{\"o}wen, H. {\em Phys. Rev. E} {\bf 1998}, {\em
  58}, 3135--3144.

\bibitem{Lik07}
Likos, C.~N.; Mladek, B.~M.; Gottwald, G.; Kahl, G. {\em J. Chem. Phys.} {\bf 2007}, {\em
  126}, 224502.


\end{thebibliography}
\end{document}